\documentclass[10pt,draft]{dis03}
\usepackage{epsf,amsmath,axodraw}

\def\beq{\begin{equation}}
\def\eeq{\end{equation}}
\def\beqa{\begin{eqnarray}}
\def\eeqa{\end{eqnarray}}

\begin{document}

\title{TWO-LOOP AND N-LOOP EIKONAL VERTEX CORRECTIONS
\thanks{The author's research has been supported 
by a Marie Curie Fellowship 
of the European Community programme ``Improving Human Research Potential'' 
under contract number HPMF-CT-2001-01221.}}

\author{NIKOLAOS KIDONAKIS \\
Cavendish Laboratory, University of Cambridge\\
Cambridge CB3 0HE, England\\
E-mail: kidonaki@hep.phy.cam.ac.uk\\ }

\maketitle

\begin{abstract}
\noindent I present calculations of two-loop vertex corrections 
with massive and massless partons in the eikonal approximation. 
I show that the $n$-loop result for the UV poles can be given in terms of the 
one-loop calculation.
\end{abstract}

\section{Introduction} 

The eikonal approximation is valid for emission of soft gluons.
The approximation simplifies the usual Feynman rules for the quark
propagator and quark-gluon vertex as follows:
\beq
{\bar u}(p) (-i \gamma^{\mu}) \frac{i (p\!\!/+k\!\!/+m)}{(p+k)^2
-m^2+i\epsilon} \rightarrow {\bar u}(p) \gamma^{\mu}
\frac{p\!\!/+m}{2p\cdot k+i\epsilon}
={\bar u}(p) \frac{v^{\mu}}{ v\cdot k+i\epsilon}
\eeq
with $k\rightarrow 0$ the gluon momentum, 
$p$ the quark momentum after emission of
the gluon, $v$ a dimensionless vector $v \propto p$, and I have omitted 
overall factors of $g_s \, T^c_F$ with $g_s^2=4\pi \alpha_s$
and $T^c_F$ the generators of SU(3) in the fundamental representation.

The eikonal approximation  has numerous phenomenological applications 
in QCD, including threshold resummations for a variety 
of QCD processes \cite{KS,KOS,LOS,LSV,NK,NKNNLO}.
In these applications  we are mainly interested in the ultraviolet (UV) 
pole structure (in dimensional regularization)
of one-loop, two-loop, and higher-loop eikonal vertex corrections.
In this talk, I discuss explicit calculations of 
one-loop and two-loop eikonal vertex
corrections for diagrams with massive and massless partons and show that the 
$n$-loop UV poles are given simply in terms of the
one-loop result \cite{NK2nloop}.

\section{One-loop calculations}
Let us denote by $\omega_{ij}^{(n)}$ the kinematics, color-independent, part
of the $n$-loop correction to the eikonal vertex with lines $i$ and $j$.
\begin{figure}[htb]
\begin{center}
\begin{picture}(120,120)(0,0)
\Vertex(0,50){5}
\ArrowLine(0,50)(60,80)
\ArrowLine(60,80)(100,100)
\Vertex(60,80){2}  
\Gluon(60,80)(60,20){2}{8}
\Text(25,78)[c]{$p_i+k$}
\Text(88,100)[c]{$p_i$}
\LongArrow(70,55)(70,45)
\Text(80,50)[c]{$k$}
\ArrowLine(0,50)(60,20)
\ArrowLine(60,20)(100,0)
\Vertex(60,20){2}
\Text(25,20)[c]{$p_j-k$}
\Text(88,0)[c]{$p_j$}
\end{picture}
\end{center}
\caption{\label{oneloop}  One-loop eikonal vertex correction diagram}
\end{figure}
At one loop, the expression for $\omega_{ij}^{(1)}$ (see fig. 1) in axial
gauge is
\beq
\omega_{ij}^{(1)}(v_{i},v_{j})\equiv
{g}_{s}^2\int\frac{d^D k}{(2\pi)^D}\frac{(-i)}{k^2+i\epsilon}
N^{\mu \nu}(k) \frac{\Delta_{i} \: v_{i}^{\mu}}
{\delta_{i} v_{i} \cdot k+i\epsilon}
\frac{\Delta_{j} \:v_{j}^{\nu}} {\delta_{j}v_{j} \cdot k+i\epsilon}
\eeq
with $\delta=+1(-1)$ when 
$k$ flows in the same (opposite) direction as $v$, and 
\beq
N^{\mu \nu}(k)=g^{\mu \nu}-\frac{n^{\mu}k^{\nu}+k^{\mu}n^{\nu}}
{n \cdot k}+n^2 \frac{k^{\mu}k^{\nu}}{(n\cdot k)^2} \, ,
\label{Nmunu}
\eeq
where $n$ is the axial gauge vector.
$\Delta=+1(-1)$ for a quark (antiquark) eikonal line, while
for a gluon eikonal line $\Delta=+i(-i)$ for a gluon located
above (below) the eikonal line.

Now let $I_{l}^{(1)}$ denote the contribution to $\omega_{ij}^{(1)}$ 
from the $l$-th term in the gluon propagator $N^{\mu \nu}(k)$ 
(i.e. $I_{1}^{(1)}$ is the contribution from the $g^{\mu \nu}$ term,
$I_{2}^{(1)}$ from the $n^{\mu}k^{\nu}$ terms, and
$I_{3}^{(1)}$ from the $k^{\mu}k^{\nu}$ term).
In dimensional regularization with $\varepsilon=4-D$, 
the UV poles in $\omega_{ij}^{(1)}$ for the case of massive quarks, 
with mass $m$, are given by \cite{KS}
\beq
I_1^{(1)\,\rm{UV}}
=\frac{\alpha_s}{\pi}\frac{1}{\varepsilon} L_{\beta} \, , \quad
I_2^{(1)\,\rm{UV}}=\frac{\alpha_s}{\pi}\frac{1}{\varepsilon} 
(L_i+L_j) \, , \quad
I_3^{(1)\,\rm{UV}}=-\frac{\alpha_s}{\pi}\frac{1}{\varepsilon} \, ,
\eeq
and thus 
\beq
\omega_{ij}^{(1)\; {\rm UV}}=
{\cal S}_{ij}^{(1)} \left[I_1^{(1)\,\rm{UV}}
+I_2^{(1)\,\rm{UV}}+I_3^{(1)\,\rm{UV}}\right]
={\cal S}_{ij}^{(1)} \, \frac{\alpha_{s}}{\pi\varepsilon}
\left[L_{\beta} + L_i + L_j -1 \right]
\nonumber
\eeq
with ${\cal S}_{ij}^{(1)}=\Delta_i \Delta_j \delta_i \delta_j$ 
an overall sign and 
\beq
L_{\beta}=\frac{1-2m^2/s}{\beta}
\left[\ln\left(\frac{1-\beta}{1+\beta}\right)+\pi i\right]
\eeq 
with $\beta=\sqrt{1-4m^2/s}$. The functions $L_i$ and $L_j$ depend
on the axial gauge vector $n$ and are cancelled when we include
the heavy-quark self energies.

When $v_i$ refers to a massive quark and $v_j$ to a massless quark 
we have \cite{KS}
\beqa
I_1^{(1)\,\rm{UV}}&=&\frac{\alpha_s}{2\pi}
\left\{\frac{2}{\varepsilon^2}-\frac{1}{\varepsilon}
\left[\gamma_E+\ln\left(\frac{v_{ij}^2 s}{2m^2}\right)-\ln(4\pi) \right]
\right\}\, , 
\nonumber\\
I_2^{(1)\,\rm{UV}}&=&\frac{\alpha_s}{2\pi}
\left\{-\frac{2}{\varepsilon^2}+\frac{1}{\varepsilon}
\left[2L_i+\gamma_E+\ln \nu_j-\ln(4\pi)\right]
\right\} \, ,
\nonumber \\  
I_3^{(1)\,\rm{UV}}&=&-\frac{\alpha_s}{\pi}\frac{1}{\varepsilon} \, ,
\eeqa
where
$\nu_a=(v_a \cdot n)^2/|n|^2$, 
$v_{ij}=v_i \cdot v_j$, and $\gamma_E$ is the Euler constant. 
Note that the double poles cancel
in the sum over the $I^{(1)}$'s  and we get
\beq       
\omega_{ij}^{(1)\; \rm{UV}}(v_{i},v_{j})=
{\cal S}_{ij}^{(1)} \, \frac{\alpha_{s}}{\pi\varepsilon}
\left[-\frac{1}{2}\ln\left(\frac{v_{ij}^2s}{2m^2}\right) + L_i 
+\frac{1}{2}\ln \nu_{j} -1\right].
\eeq

Finally, when both $v_i$ and $v_j$ refer to massless quarks we have
\cite{BotSt,KS}
\beqa
I_1^{(1)\,\rm{UV}}&=&\frac{\alpha_s}{\pi}
\left\{\frac{2}{\varepsilon^2}-\frac{1}{\varepsilon}
\left[\gamma_E+\ln\left(\delta_i\delta_j\; \frac{v_{ij}}{2}\right)
-\ln(4\pi) \right]\right\} \, ,
\nonumber\\ 
I_2^{(1)\,\rm{UV}}&=&\frac{\alpha_s}{\pi}
\left\{-\frac{2}{\varepsilon^2}+\frac{1}{\varepsilon}
\left[\gamma_E+\frac{1}{2}\ln(\nu_i \nu_j)-\ln(4\pi)\right]\right\} \, ,
\nonumber \\
I_3^{(1)\,\rm{UV}}&=&-\frac{\alpha_s}{\pi}\frac{1}{\varepsilon} \, .
\eeqa
Again, we note that the double poles cancel in the sum over the $I^{(1)}$'s
and we get
\beq
\omega_{ij}^{(1)\; \rm{UV}}(v_{i},v_{j})=
{\cal S}_{ij}^{(1)} \, \frac{\alpha_{s}}{\pi\varepsilon}
\left[-\ln\left(\delta_{i} \, \delta_{j} \, 
\frac{v_{ij}}{2}\right)+\frac{1}{2}\ln(\nu_{i}\nu_{j})-1\right].
\eeq

\section{Two-loop and $n$-loop calculations}

For the two-loop diagram in fig. 2 we have

\begin{figure}[htb]
\begin{center}
\begin{picture}(120,120)(0,0)
\Vertex(0,60){5}
\ArrowLine(0,60)(60,85)
\ArrowLine(60,85)(100,102)
\ArrowLine(100,102)(120,110)
\Vertex(60,85){2}
\Vertex(60,35){2}  
\Gluon(60,85)(60,35){2}{8}
\Text(15,90)[c]{$p_i+k_1+k_2$}
\Text(75,105)[c]{$p_i+k_1$}
\Text(110,115)[c]{$p_i$}
\LongArrow(70,65)(70,55)
\Text(80,60)[c]{$k_2$}
\ArrowLine(0,60)(60,35)
\ArrowLine(60,35)(100,18)
\ArrowLine(100,18)(120,10)
\Vertex(100,102){2}
\Vertex(100,18){2}
\Gluon(100,102)(100,18){2}{8}
\Text(15,30)[c]{$p_j-k_1-k_2$}
\Text(75,15)[c]{$p_j-k_1$}
\Text(110,5)[c]{$p_j$}
\LongArrow(110,65)(110,55)
\Text(120,60)[c]{$k_1$}
\end{picture}
\end{center}
\caption{\label{twoloop}  Two-loop eikonal vertex correction diagram}
\end{figure}

\beqa
\omega_{ij}^{(2)}(v_{i},v_{j})&=&
{g}_{s}^4 \int\frac{d^D k_1}{(2\pi)^D} \, \frac{(-i)}{k_1^2+i\epsilon} \,
N^{\mu \nu}(k_1)
\frac{\Delta_{1i} \: v_{i}^{\mu}}
{\delta_{1i} v_{i} \cdot k_1+i\epsilon} \;
\frac{\Delta_{1j} \:v_{j}^{\nu}}
{\delta_{1j}v_{j} \cdot k_1+i\epsilon}
\nonumber \\ && \hspace{-31mm} \times \,
\int\frac{d^D k_2}{(2\pi)^D} \, \frac{(-i)}{k_2^2+i\epsilon} \,
N^{\rho \sigma}(k_2)
\frac{\Delta_{2i} \: v_{i}^{\rho}}
{\delta_{2i} v_{i} \cdot (k_1+k_2)+i\epsilon} \;
\frac{\Delta_{2j} \:v_{j}^{\sigma}}
{\delta_{2j}v_{j} \cdot (k_1+k_2)+i\epsilon}.
\eeqa

Let $I_{kl}^{(2)}$ denote the contribution to $\omega_{ij}^{(2)}$ from the 
product of the $k$-th term in the axial-gauge gluon propagator 
$N^{\mu \nu}(k_1)$ with the $l$-th term in $N^{\rho \sigma}(k_2)$.
Then we can rewrite
\beq
\omega^{(2)}_{ij}(v_{i},v_{j})= 
{\cal S}_{ij}^{(2)}
\sum_{k,l=1,2,3} I_{kl}^{(2)}(v_i, v_j) \, .
\eeq
When both partons are massive, an explicit calculation of the two-loop
diagram gives the following results for the leading UV ($1/\varepsilon^2$)
poles at two loops: 
\beqa
I_{11}^{(2), \, {\rm UV}}&=&\frac{\alpha_s^2}{\pi^2}
\frac{1}{\varepsilon^2} L_{\beta}^2\, , \quad
I_{12}^{(2), \, {\rm UV}}=\frac{\alpha_s^2}{\pi^2}
\frac{1}{\varepsilon^2} \, L_{\beta} \left(L_i+L_j\right)
=I_{21}^{(2), \, {\rm UV}}
\nonumber \\
I_{22}^{(2), \, {\rm UV}}&=&\frac{\alpha_s^2}{\pi^2}
\frac{1}{\varepsilon^2} \left(L_i+L_j\right)^2\, , \quad
I_{13}^{(2), \, {\rm UV}}=-\frac{\alpha_s^2}{\pi^2}
\frac{1}{\varepsilon^2} L_{\beta}
=I_{31}^{(2), \, {\rm UV}}
\nonumber \\
I_{23}^{(2), \, {\rm UV}}&=&-\frac{\alpha_s^2}{\pi^2}
\frac{1}{\varepsilon^2} \left(L_i+L_j\right)
=I_{32}^{(2), \, {\rm UV}} \, , \quad
I_{33}^{(2), \, {\rm UV}}=\frac{\alpha_s^2}{\pi^2}
\frac{1}{\varepsilon^2} \, .
\eeqa
Then
\beqa
&&\omega_{ij}^{(2)\; \rm{UV}}(v_{i},v_{j})=
{\cal S}_{ij}^{(2)} \, \frac{\alpha_{s}^2}{\pi^2}
\frac{1}{\varepsilon^2}
\left[L_{\beta}+L_i + L_j -1 \right]^2
+{\cal O} \left(\frac{1}{\varepsilon}\right) \, ,
\label{omega2heavy}
\eeqa
where ${\cal S}_{ij}^{(2)}=\Delta_{1i} \Delta_{1j} \Delta_{2i} \Delta_{2j}
\delta_{1i} \delta_{1j} \delta_{2i} \delta_{2j}$.
The calculation of the $1/\varepsilon$ terms is given in \cite{NK2nloop}.

We now note that the leading two-loop UV poles are simply the 
square of the one-loop
result since $I_{mn}^{(2), \, {\rm UV}}=I_{m}^{(1), \, {\rm UV}}
I_{n}^{(1), \, {\rm UV}}$.
Similar results hold for the cases when one or both partons are massless.
For example, the leading poles in $I_{23}^{(2), \, {\rm UV}}$ 
when one of the partons is massless and the other is massive are 
$(\alpha_s^2/\pi^2)(1/\varepsilon^3)$.

We noted that the coefficient of the leading UV pole
in $\omega_{ij}^{(2)}$ is simply the square
of the coefficient of the leading UV pole in $\omega_{ij}^{(1)}$.
We can show by induction that this generalizes to $n$ loops
\cite{NK2nloop}. 
Thus the leading UV pole in the $n$-loop corrections for massive partons,
$\omega_{ij}^{(n)}$, is $(\alpha_{s}/\pi)^n
(1/\varepsilon^n)\left[L_{\beta}+L_i + L_j -1 \right]^n$.
Furthermore a similar structure holds for  
non-leading UV poles \cite{NK2nloop}.


\begin{thebibliography}{0}

\bibitem{KS}
N. Kidonakis and G. Sterman, Phys. Lett. B {\bf 387}, 867 (1996);
Nucl. Phys. {\bf B505}, 321 (1997).

\bibitem{KOS}
N. Kidonakis, G. Oderda, and G. Sterman,
Nucl. Phys. {\bf B525}, 299 (1998);
Nucl. Phys. {\bf B531}, 365 (1998).

\bibitem{LOS}
E. Laenen, G. Oderda, and G. Sterman, 
Phys. Lett. B {\bf 438}, 173 (1998).

\bibitem{LSV}
E. Laenen, G. Sterman, and W. Vogelsang,
Phys. Rev. D {\bf 63}, 114018 (2001). 

\bibitem{NK} 
N. Kidonakis, Phys. Rev. D {\bf 64}, 014009 (2001);
Int. J. Mod. Phys. A {\bf 15}, 1245 (2000).

\bibitem{NKNNLO}
N. Kidonakis, Cavendish-HEP-03/02, hep-ph/0303186; in {\it DIS03},
hep-ph/0306125.

\bibitem{NK2nloop}
N. Kidonakis, hep-ph/0208056; in {\it DPF2002} hep-ph/0207142.

\bibitem{BotSt} 
J. Botts and G. Sterman, Nucl. Phys. {\bf B325}, 62 (1989). 

\end{thebibliography}
\end{document}